\documentclass[11pt]{article}
\usepackage[margin=1in]{geometry}
\usepackage{amsmath,amssymb,graphicx,booktabs,xcolor}
\usepackage{tikz}\usetikzlibrary{positioning,arrows.meta}
\usepackage[numbers,sort&compress]{natbib}
\usepackage[colorlinks=true,linkcolor=blue,citecolor=blue,urlcolor=blue]{hyperref}
\title{Matching two-dimensional statistics does not certify three-dimensional transport in battery electrodes}
\author{Maziar Raissi$^{1,*}$ \and Devesh Upadhyay$^{2}$\\[4pt]
\small $^{1}$Department of Mathematics, University of California, Riverside, Riverside, California 92521, USA\\
\small $^{2}$Saab, Inc., East Syracuse, New York 13057, USA\\
\small $^{*}$Corresponding author: maziar.raissi@ucr.edu}
\date{}
\begin{document}
\maketitle

\begin{abstract}
Three-dimensional battery electrode structures reconstructed from two-dimensional micrographs are routinely used to predict transport,
assuming that matching the measured 2D statistics reproduces the relevant 3D properties. We test this premise on X-ray
tomograms of three all-solid-state cathodes. With a differentiable conductivity solver, we optimise voxel structures that match the real
electrode in phase fractions, boundary density, two- and three-point correlations and lineal-path functions, on all section planes, slab
by slab and layer by layer, within measurement noise. Every sample has statistical twins conducting 10--88-fold less than the
real electrode, robust to $5^3$ majority filtering, whereas the most conductive are at most 3.3-fold above it. The twins rearrange rather than remove material,
severing or funnelling through-plane paths. A classifier on 2D sections detects them, as it does current 2D-to-3D generators: the
information is in the images but not in the statistics. Matching standard 2D statistics therefore does not certify 3D transport.
\end{abstract}

\section{Introduction}
Battery electrode performance is governed by transport through a three-dimensional (3D) composite of active material, electrolyte and
pores. Full 3D imaging by X-ray or focused-ion-beam tomography is expensive and limited in field of view, so the field increasingly
reconstructs 3D structures from 2D micrographs: stochastic geometry models calibrated on 2D sections~\cite{furat2025}, generative
adversarial networks trained on slices~\cite{kench2021}, and statistical reconstruction from correlation functions~\cite{yeong1998,jiao2007,bostanabad2018}, now available as large
libraries of generated 3D structures~\cite{kench2022}.
Effective transport is then simulated on the reconstruction and reported as a property of the material~\cite{dufter2026,cooper2016}, or predicted directly from geometric descriptors
through fitted laws~\cite{stenzel2016,neumann2020}.

This workflow assumes that the 2D statistics identify the 3D property. Classical stereology says when that holds: volume fractions and
surface densities are fixed by planar sections, connectivity is not~\cite{torquato2002}. Effective conductivity depends on
connectivity, so it is unclear how much of it a set of 2D statistics determines. A reconstruction that reproduces the statistics shows
that they are consistent with one 3D structure; it says nothing about the other structures that fit equally well.

Non-uniqueness of reconstructions is well known. Different structures can share a two-point correlation function, and the size of
this degenerate set has been quantified~\cite{gommes2012}; lineal-path and higher-order functions were introduced partly to shrink
it~\cite{lu1992,jiao2007}. In battery electrodes, tortuosity factors estimated from 2D information or from different methods disagree
substantially and depend on direction~\cite{ebner2014,usseglio2018}. What has been missing is the consequence for transport: how
far apart in effective conductivity can structures be that match a realistic set of 2D statistics within measurement noise, and
what distinguishes them? Answering this requires searching the degenerate set for its transport extremes rather than sampling it.

We pose the question as one of partial identification~\cite{manski2003}: given the named 2D statistics of a real electrode, measured with
their natural noise, what range of through-thickness effective conductivity $M=\sigma_{\mathrm{eff}}/\sigma_0$ is consistent with them?
We answer it constructively, by optimising directly over voxel fields under the statistical constraints. Every structure we report is a
verified witness: it satisfies all constraints, and its conductivity comes from a converged solve. A local search cannot certify the
global extremes, so the ranges we report are inner bounds, that is, sub-intervals of the identified set; the true range is at least as wide.

Our contributions are: (i) an audit showing that current 2D-to-3D generators are detected by a calibrated classifier two-sample test
(C2ST)~\cite{lopezpaz2017}; (ii) a differentiable bound engine for effective conductivity over voxel fields, with exact gradients and no
adjoint solve; (iii) resolution-robust witnesses on three real cathodes that match a hierarchy of stereological statistics while
conducting at least 33-fold less than the real electrode in every sample (10--88-fold across runs); and (iv) the mechanism: severed or funnelled interior connectivity,
which the named statistics do not see.

\section{Results}
\subsection{Current 2D-to-3D generators are detectable}
\begin{figure}[t]\centering
\includegraphics[width=\textwidth]{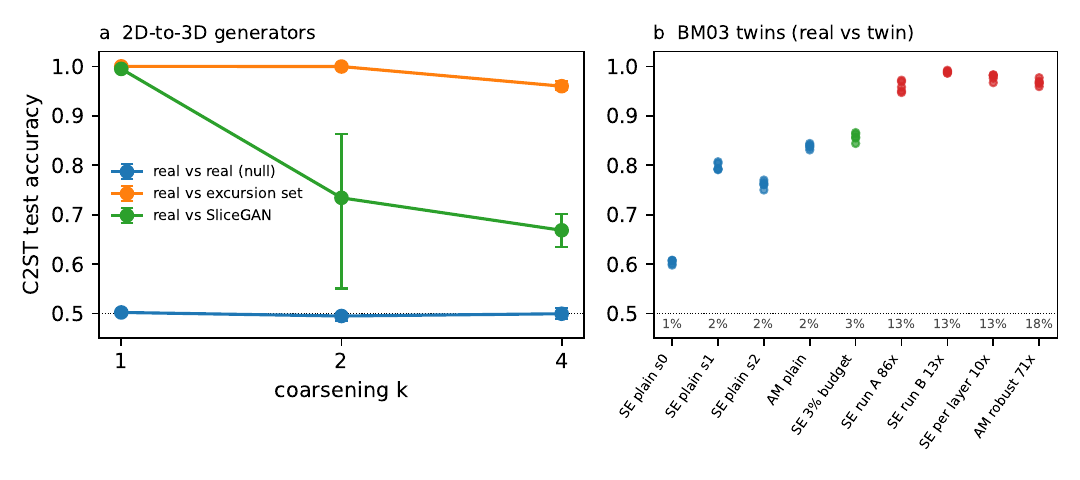}
\caption{Classifier two-sample tests (five seeds; 0.5 = indistinguishable). (a) Real BM03 sections versus 2D-to-3D generators,
after $k\times k$ majority-vote coarsening; the real-versus-real null stays at 0.50. (b) Real BM03 block versus minimum-conductivity
twins (Table~\ref{tab:witness}): plain-objective twins that are not resolution-robust (blue), the twin limited to 3\% of voxels (green)
and the resolution-robust twins (red). Numbers above the axis: fraction of voxels changed; s0--s2: three seeds of the plain objective.}
\label{fig:audit}\end{figure}

On BM03 sections ($128^2$ pixels), a CNN classifier trained on fixed train, validation and test pools with early stopping gives 0.50 test
accuracy for real versus real (the calibrated null), but 1.00 for real versus the calibrated excursion-set model~\cite{furat2025} and
1.00/0.995 for real versus SliceGAN~\cite{kench2021} (two checkpoints). After majority-vote coarsening by $k=2$ and $4$, SliceGAN drops to
0.73 and 0.67 while the excursion model stays at 1.00 and 0.96. SliceGAN thus differs mainly in fine-scale texture, the excursion model in
morphology (Fig.~\ref{fig:audit}a).

\subsection{A differentiable bound engine}
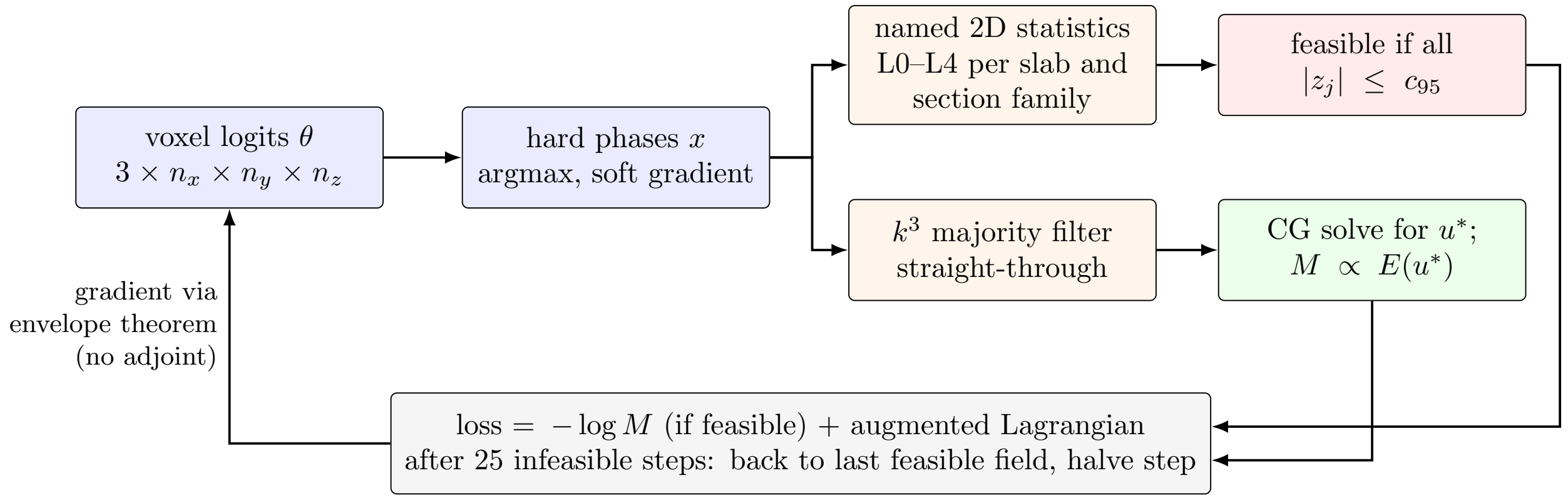
\begin{figure}[t]\centering
\resizebox{\textwidth}{!}{\begin{tikzpicture}[font=\small,
  box/.style={draw, rounded corners=2pt, align=center, minimum height=12mm, text width=34mm, fill=#1!8},
  box/.default=gray, arr/.style={-{Latex[length=2mm]}, thick}]
\node[box=blue]   (theta) at (0,0)      {voxel logits $\theta$\\$3\times n_x\times n_y\times n_z$};
\node[box=blue]   (hard)  at (4.6,0)    {hard phases $x$\\argmax, soft gradient};
\node[box=orange] (stats) at (9.2,1.1)  {named 2D statistics\\L0--L4 per slab and\\section family};
\node[box=orange] (filt)  at (9.2,-1.1) {$k^3$ majority filter\\straight-through};
\node[box=red]    (feas)  at (13.6,1.1) {feasible if all\\$|z_j|\le c_{95}$};
\node[box=green]  (solve) at (13.6,-1.1){CG solve for $u^*$;\\$M\propto E(u^*)$};
\node[box, text width=95mm] (loss) at (6.8,-3.4) {loss $=-\log M$ (if feasible) $+$ augmented Lagrangian\\
  after 25 infeasible steps: back to last feasible field, halve step};
\draw[arr] (theta) -- (hard);
\draw[arr] (hard.east) -- ++(0.5,0) |- (stats.west);
\draw[arr] (hard.east) -- ++(0.5,0) |- (filt.west);
\draw[arr] (stats) -- (feas);
\draw[arr] (filt) -- (solve);
\draw[arr] (feas.east) -- ++(0.4,0) |- ([yshift=2mm]loss.east);
\draw[arr] (solve.south) |- ([yshift=-2mm]loss.east);
\draw[arr] (loss.west) -| node[pos=0.75, left, align=right, font=\footnotesize]{gradient via\\envelope theorem\\(no adjoint)} (theta.south);
\end{tikzpicture}}
\caption{Bound engine. The voxel field is optimised directly. Forward: hard phases, the named 2D statistics on every 32-voxel slab and
all three section families, and the through-thickness conductivity $M$ of the majority-filtered field from a converged conjugate-gradient
solve. Backward: straight-through gradients through the phase choice and the filter; $\mathrm{d}M/\mathrm{d}c$ follows from the envelope
theorem at the stationary potential, so no adjoint solve is needed. Only feasible fields are recorded as witnesses. Shown for
minimisation; maximisation uses $+\log M$.}
\label{fig:engine}\end{figure}

Figure~\ref{fig:engine} summarises the method. The solver reproduces a reference finite-volume solver on the BM03 tomogram (0.14186 and 0.26878 for the two
conducting phases), and its gradient agrees with finite differences to about seven digits. One $128^3$ solve with backpropagation takes
about 0.5\,s on one GPU.

\subsection{Statistical twins with far lower conductivity}
\begin{table}[t]\centering\small
\caption{Minimum-conductivity witnesses. $M$ of the witness after $k^3$ majority filtering and the reduction factor against the real block
filtered with the same kernel. ``Robust'': $M$ of the $k=3$ filtered field was the objective; ``plain'': unfiltered objective. SE: solid
electrolyte (ionic); AM: active material (electronic). Perc.: fraction of the phase connected top to bottom, after $k=3$ (real in brackets).
All witnesses satisfy every constraint of the L4 hierarchy with 8-layer depth bins.}
\label{tab:witness}
\begin{tabular}{lll rr rr rr l}\toprule
& & & \multicolumn{2}{c}{$k=1$} & \multicolumn{2}{c}{$k=3$} & \multicolumn{2}{c}{$k=5$} & \\
\cmidrule(lr){4-5}\cmidrule(lr){6-7}\cmidrule(lr){8-9}
sample & phase & objective & $M$ & $\times$ & $M$ & $\times$ & $M$ & $\times$ & perc. \\\midrule
BM01 & SE & plain & 0.0007 & 81 & 0.0019 & 33 & 0.0012 & 48 & 0.00 (0.98) \\
BM03 & SE & robust, run A & 0.0022 & 88 & 0.0025 & 86 & 0.0026 & 86 & 0.98 (0.99) \\
BM03 & SE & robust, run B$^a$ & 0.0111 & 18 & 0.0164 & 13 & 0.0190 & 12 & 0.78 (0.99) \\
BM03 & SE & robust, per layer$^b$ & 0.0163 & 12 & 0.0212 & 10 & 0.0224 & 10 & 0.99 (0.99) \\
BM03 & SE & robust, 3\% budget$^c$ & 0.1286 & 1.5 & 0.1374 & 1.6 & 0.1432 & 1.6 & 0.99 (0.99) \\
BM03 & AM & robust & 0.0036 & 65 & 0.0040 & 71 & 0.0043 & 77 & 1.00 (1.00) \\
BM10 & SE & robust & 0.0005 & 59 & 0.0005 & 72 & 0.0005 & 78 & 0.00 (0.82) \\
\bottomrule\end{tabular}
\\[2pt]\footnotesize $^a$Independent start; continued with a 100$\times$ lower step-size floor until the step size collapsed.
$^b$Phase fractions of every single layer constrained (about $\pm0.017$) instead of 8-layer bins; stopped by the step-size floor while $M$ was still falling; continuing with a 100$\times$ lower step-size floor gave no further reduction. $^c$At most 3\% of voxels may differ from the real block.\end{table}
Table~\ref{tab:witness} lists the witnesses. Every sample admits a twin whose electrolyte conductivity is at least 33-fold below the
real block's after filtering with the same kernel: BM01 33--81$\times$, BM10 59--78$\times$, BM03 86--88$\times$. For BM01 the best
witness came from the unfiltered objective and still holds up after filtering; the robust-objective BM01 runs stopped at 2--4$\times$ (one ended at the step-size floor, one still improving).
An independent replicate of the BM03 run from a perturbed start
stalled at $M=0.0164$ ($13\times$ lower; 12--18$\times$ across $k$; run B in Table~\ref{tab:witness}) even when continued with a
100$\times$ lower step-size floor, so the factor depends on the search path. The mechanism does not: both replicates cut the interior (below). Constraining the phase fractions of every single layer instead of 8-layer bins does not close the gap: that twin is
10--12$\times$ below the real block at all three filter sizes, with percolation intact (0.99) (Table~\ref{tab:witness}). Its interior is
not fully severed: three of four quadrants conduct nothing, the rest about $10\times$ less than the real block. Active material (electronic conduction, BM03): $M=0.0036$/0.0040/0.0043 at $k=1,3,5$ against
0.233/0.284/0.329, 65--77$\times$ lower, with the same interior-cut mechanism (every lateral quadrant 0.0000).

All ranges are inner bounds. Long runs kept improving after plateaus of thousands of steps (BM03 fell 40-fold after 2{,}900 slow steps; Supplementary Fig.~S1), so
stalling of a local search is no evidence of a global extreme. The same holds for the upper end: short unfiltered maximisations (3{,}000 steps) gave witnesses 1.3--1.9$\times$ above the real
block, and longer resolution-robust runs (9{,}000 steps, $3^3$-filtered objective; Table~\ref{tab:max}) raised this to 1.3--3.3$\times$.
The most conductive twins of the two well-connected blocks (BM03, real percolating fraction ${\ge}0.98$) stay within 1.3--1.4$\times$; the
larger gains are for BM01 and BM10, whose real electrolyte is less connected (percolating fraction 0.97 and 0.82) and which the twins
complete (0.98 in both). The witnessed range is thus strongly asymmetric: the named statistics permit structures 10--88$\times$ less conductive,
but at most $3.3\times$ more conductive.

\begin{table}[t]\centering\small
\caption{Most conductive verified twins (resolution-robust maximisation, 9{,}000 steps), evaluated after $k^3$ majority filtering, with the
real block filtered the same way. Ratio = twin / real. Perc.: percolating fraction of the conducting phase at $k=1$ (real in brackets).}
\label{tab:max}
\begin{tabular}{llccccccc}\toprule
 & & \multicolumn{2}{c}{$k=1$} & \multicolumn{2}{c}{$k=3$} & \multicolumn{2}{c}{$k=5$} & \\
sample & phase & $M$ (real) & ratio & $M$ (real) & ratio & $M$ (real) & ratio & perc. \\\midrule
BM01 & SE & 0.114 (0.058) & 2.0 & 0.152 (0.063) & 2.4 & 0.125 (0.060) & 2.1 & 0.98 (0.97) \\
BM03 & SE & 0.268 (0.198) & 1.35 & 0.307 (0.215) & 1.43 & 0.305 (0.224) & 1.36 & 0.99 (0.99) \\
BM03 & AM & 0.314 (0.233) & 1.35 & 0.397 (0.284) & 1.40 & 0.436 (0.329) & 1.32 & 1.00 (1.00) \\
BM10 & SE & 0.087 (0.032) & 2.7 & 0.127 (0.039) & 3.3 & 0.101 (0.041) & 2.4 & 0.98 (0.82) \\\bottomrule
\end{tabular}\end{table}

\subsection{Mechanism: severed interior connectivity}
In the BM03 witness the conductivity of every interior sub-block is zero to solver precision (the inner block after removing 64
voxels from each lateral face, and three of the four lateral quadrants), against 0.22--0.38 in the real electrode. The residual current
runs along the no-flux lateral faces: at the narrowest layer, 95\% of it passes within 32 voxels of the boundary (real: 27\%), and 533
faces carry 90\% of it (real: at least 9{,}779). The top-to-bottom percolating fraction of the electrolyte stays high (0.977, real 0.985)
only because the cluster connects through the rim of the block. A percolating fraction measured on a finite block is therefore not a
safe proxy for transport. It also means that the reported $M$ of this twin is carried by the block boundary: in a laterally larger
or periodic domain, where no rim exists, the interior cut would leave it lower still, so the finite block understates the reduction. The replicate and the active-material twin show the same pattern: the inner block (64 voxels removed from each face) conducts 0.0000,
as do three of four quadrants (replicate) and all four (active material). The BM01 twin behaves alike (every sub-block 0.0000, real 0.014--0.076). For BM10 the test cannot
discriminate: the real block's own sub-blocks conduct nothing, because its few electrolyte paths need more lateral room than a quadrant
offers. There the cut is shown by percolation instead (twin 0.0, real 0.82). 

\begin{figure}[t]\centering\includegraphics[width=\linewidth]{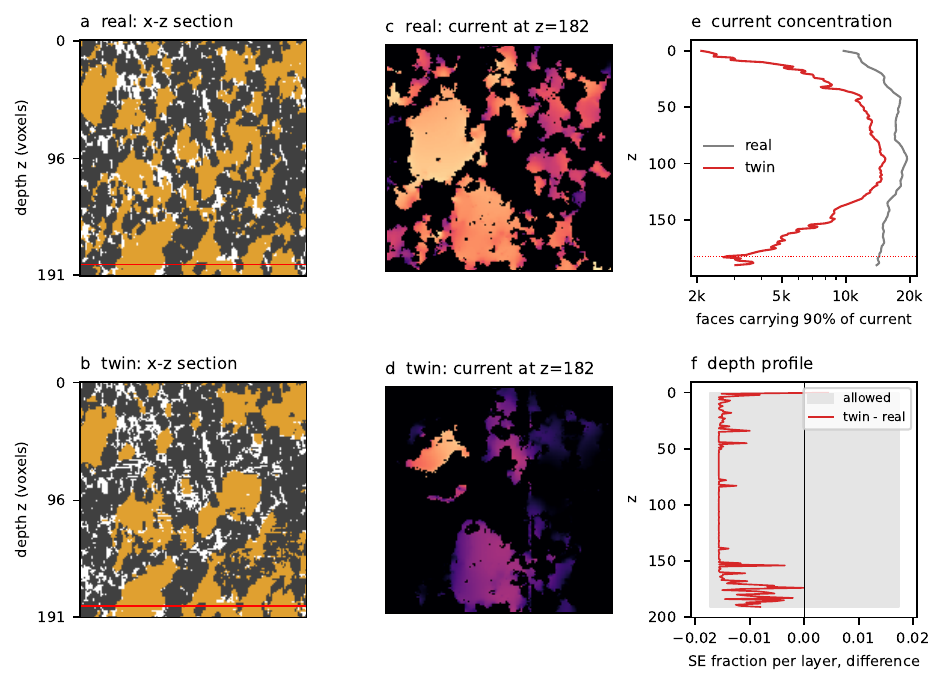}
\caption{Mechanism of the per-layer twin (BM03, electrolyte). (a,b) $x$-$z$ sections of the real block and the twin (electrolyte orange,
active material grey, pore white); the red line marks the layer of strongest current concentration. (c,d) Through-plane current at that
layer, log scale over four decades, same colour range. (e) Number of voxel faces that carry 90\% of the current in each layer: the twin
funnels its current through up to $5\times$ fewer faces. (f) Twin minus real electrolyte fraction per layer with the allowed band.}
\label{fig:mech}\end{figure}

Figure~\ref{fig:mech} shows how the per-layer twin works. It keeps every layer's electrolyte fraction inside the allowed band, but uses
the band in one direction: on average it holds 1.4\% less electrolyte per layer. That deficit alone does not explain the drop. Taking the
same number of electrolyte voxels out of each layer of the real block, chosen at random or from the particle surfaces, lowers $M$ by only
7--8\% (0.198 to 0.182--0.185), against $12\times$ for the twin. What matters is where the electrolyte is removed: the twin funnels the
current of its deepest layers through up to $5\times$ fewer faces than the real block (Fig.~\ref{fig:mech}e).

\subsection{Are the twins detectable?}
A CNN classifier trained to tell 2D sections of a twin from the same sections of the real block (Methods) separates them easily
when the twin is resolution-robust: test accuracy 0.95--0.97 for the $86\times$ electrolyte twin (13\% of voxels changed) and
0.96--0.98 for the active-material twin (18\%), 0.97--0.98 for the per-layer twin (13\%) and 0.99 for run B (13\%), over five seeds. Earlier, non-robust twins that changed 1.3--1.9\% of voxels reached only
0.60--0.84, but their cuts healed under $3^3$ filtering (Fig.~\ref{fig:audit}b). Capping the edit at 3\% of voxels gives a twin that is
still detected (0.84--0.87) and is only $1.6\times$ less conductive. Large, resolution-robust reductions thus require editing a
sizeable fraction of the volume, and these edits leave traces in the 2D sections that a classifier can find. The twins match every named 2D statistic; they are
not indistinguishable from the real electrode. The point is therefore not that 2D statistics are fooled by an invisible change, but that
the statistics in standard use leave the transport-controlling connectivity unconstrained.

\section{Discussion}
\emph{What the named statistics leave open.} On three cathodes and for both conducting phases, structures that match the full L0--L4
hierarchy within measurement noise, on all three section families, slab by slab and with the depth profile held layer by layer,
conduct one to two orders of magnitude less than the real electrode. Since all reported structures are witnesses of a local search,
these are inner bounds: the consistent range of $M$ is at least this wide. A conductivity predicted from a reconstruction that matches
these statistics therefore carries no guarantee near the real value; the agreement reported for such reconstructions is a property of
the generator, not of the statistics it matches. This is a statement about guarantees, not about typical error: on BM03 the
calibrated excursion-set model and SliceGAN give electrolyte $M$ of 0.137 and 0.116 against 0.142 for the real
tomogram (full tomogram; our calibration and SliceGAN training), errors of 4--18\%. Such agreement is what generators deliver in practice; the twins show that it is not implied
by the statistics they match, so it cannot be assumed for a new material or generator without a 3D check.

\emph{Why.} The twins do not lower the amount of electrolyte; a matched random or surface removal changes $M$ by under 10\%. They
rearrange it so that through-plane paths are cut in the interior (BM01, BM10, BM03 run A and the active-material twin) or funnelled
through a few constrictions (per-layer twin), while sections keep their phase fractions, interface density, correlations and chord
statistics. These descriptors are averages over many short-range configurations; through-plane conductivity is controlled by rare
long-range ones. This is the voxel-level version of the classical observation that planar sections fix volume and surface densities
but not connectivity~\cite{torquato2002}.

\emph{The images contain more than the statistics.} The resolution-robust twins are detected by a classifier on 2D sections. The
information that rules them out is therefore present in the micrographs, just not in the descriptors that are matched. Two practical
consequences follow. First, a reconstruction should be validated with a calibrated two-sample test on sections, which current
generators fail outright, rather than by matching a list of descriptors. Second, if 3D transport is to be inferred from 2D data,
connectivity has to enter explicitly: for example through cross-sections that resolve through-plane paths, a measured cell resistance
used as a constraint, or a limited 3D acquisition of the interior.

\emph{Limitations.} The search is local, so the reductions are lower bounds on what the statistics permit and their size varies
with the start (10--88$\times$ for BM03 electrolyte across runs and filters). The upper side also rose with longer runs (from 1.3--1.9$\times$ to 1.3--3.3$\times$), so neither end is settled; the asymmetry between them, however, persisted in every run. The noise model is
estimated from the four lateral quadrants of one tomogram, which vary less than separate electrodes or micrographs do, so our
constraints are tighter than practice requires and real-world bounds will be wider still. Blocks are finite with no-flux lateral faces, and some twins conduct only along these faces; as argued above this makes the reported reductions conservative, but bounds on periodic or larger domains were not computed. Resolution enters only through majority filtering up to $5^3$; the twins at $k=1$
are voxel structures and their manufacturability is not addressed. Finally, a twin with only 3\% of voxels changed was found to be just
$1.6\times$ less conductive, so whether small, undetectable edits can give large reductions remains open.

\bibliographystyle{unsrtnat}
\bibliography{refs}

\begin{thebibliography}{18}
\providecommand{\natexlab}[1]{#1}
\providecommand{\url}[1]{\texttt{#1}}
\expandafter\ifx\csname urlstyle\endcsname\relax
  \providecommand{\doi}[1]{doi: #1}\else
  \providecommand{\doi}{doi: \begingroup \urlstyle{rm}\Url}\fi

\bibitem[Furat et~al.(2025)Furat, Weber, Dufter, Schubert, Rekers, Luczak, Glatt, Wiegmann, Janek, Bielefeld, and Schmidt]{furat2025}
O.~Furat, S.~Weber, A.~Dufter, J.~Schubert, R.~Rekers, M.~Luczak, E.~Glatt, A.~Wiegmann, J.~Janek, A.~Bielefeld, and V.~Schmidt.
\newblock Generative adversarial framework to calibrate excursion set models for the {3D} morphology of all-solid-state battery cathodes.
\newblock \emph{Adv. Intell. Syst.}, 2025.
\newblock \doi{10.1002/aisy.202500572}.

\bibitem[Kench and Cooper(2021)]{kench2021}
S.~Kench and S.~J. Cooper.
\newblock Generating three-dimensional structures from a two-dimensional slice with generative adversarial network-based dimensionality expansion.
\newblock \emph{Nat. Mach. Intell.}, 3:\penalty0 299--305, 2021.

\bibitem[Yeong and Torquato(1998)]{yeong1998}
C.~L.~Y. Yeong and S.~Torquato.
\newblock Reconstructing random media.
\newblock \emph{Phys. Rev. E}, 57:\penalty0 495--506, 1998.

\bibitem[Jiao et~al.(2007)Jiao, Stillinger, and Torquato]{jiao2007}
Y.~Jiao, F.~H. Stillinger, and S.~Torquato.
\newblock Modeling heterogeneous materials via two-point correlation functions: Basic principles.
\newblock \emph{Phys. Rev. E}, 76:\penalty0 031110, 2007.
\newblock \doi{10.1103/PhysRevE.76.031110}.

\bibitem[Bostanabad et~al.(2018)Bostanabad, Zhang, Li, Kearney, Brinson, Apley, Liu, and Chen]{bostanabad2018}
R.~Bostanabad, Y.~Zhang, X.~Li, T.~Kearney, L.~C. Brinson, D.~W. Apley, W.~K. Liu, and W.~Chen.
\newblock Computational microstructure characterization and reconstruction: Review of the state-of-the-art techniques.
\newblock \emph{Prog. Mater. Sci.}, 95:\penalty0 1--41, 2018.
\newblock \doi{10.1016/j.pmatsci.2018.01.005}.

\bibitem[Kench et~al.(2022)Kench, Squires, Dahari, and Cooper]{kench2022}
S.~Kench, I.~Squires, A.~Dahari, and S.~J. Cooper.
\newblock {MicroLib}: A library of {3D} microstructures generated from {2D} micrographs using {SliceGAN}.
\newblock \emph{Sci. Data}, 9:\penalty0 645, 2022.
\newblock \doi{10.1038/s41597-022-01744-1}.

\bibitem[Dufter et~al.(2026)Dufter, Weber, Furat, Schubert, Rekers, Luczak, Glatt, Wiegmann, Bielefeld, and Schmidt]{dufter2026}
A.~Dufter, S.~Weber, O.~Furat, J.~Schubert, R.~Rekers, M.~Luczak, E.~Glatt, A.~Wiegmann, A.~Bielefeld, and V.~Schmidt.
\newblock Virtual materials testing of {ASSB} cathodes combining {AI}-based stochastic {3D} modeling and numerical simulations.
\newblock \emph{Mater. Today Commun.}, 2026.
\newblock \doi{10.1016/j.mtcomm.2026.115724}.

\bibitem[Cooper et~al.(2016)Cooper, Bertei, Shearing, Kilner, and Brandon]{cooper2016}
S.~J. Cooper, A.~Bertei, P.~R. Shearing, J.~A. Kilner, and N.~P. Brandon.
\newblock {TauFactor}: An open-source application for calculating tortuosity factors from tomographic data.
\newblock \emph{SoftwareX}, 5:\penalty0 203--210, 2016.
\newblock \doi{10.1016/j.softx.2016.09.002}.

\bibitem[Stenzel et~al.(2016)Stenzel, Pecho, Holzer, Neumann, and Schmidt]{stenzel2016}
O.~Stenzel, O.~Pecho, L.~Holzer, M.~Neumann, and V.~Schmidt.
\newblock Predicting effective conductivities based on geometric microstructure characteristics.
\newblock \emph{AIChE J.}, 62:\penalty0 1834--1843, 2016.
\newblock \doi{10.1002/aic.15160}.

\bibitem[Neumann et~al.(2020)Neumann, Stenzel, Willot, Holzer, and Schmidt]{neumann2020}
M.~Neumann, O.~Stenzel, F.~Willot, L.~Holzer, and V.~Schmidt.
\newblock Quantifying the influence of microstructure on effective conductivity and permeability: Virtual materials testing.
\newblock \emph{Int. J. Solids Struct.}, 184:\penalty0 211--220, 2020.
\newblock \doi{10.1016/j.ijsolstr.2019.03.028}.

\bibitem[Torquato(2002)]{torquato2002}
S.~Torquato.
\newblock \emph{Random Heterogeneous Materials: Microstructure and Macroscopic Properties}.
\newblock Springer, New York, 2002.

\bibitem[Gommes et~al.(2012)Gommes, Jiao, and Torquato]{gommes2012}
C.~J. Gommes, Y.~Jiao, and S.~Torquato.
\newblock Microstructural degeneracy associated with a two-point correlation function and its information content.
\newblock \emph{Phys. Rev. E}, 85:\penalty0 051140, 2012.
\newblock \doi{10.1103/PhysRevE.85.051140}.

\bibitem[Lu and Torquato(1992)]{lu1992}
B.~Lu and S.~Torquato.
\newblock Lineal-path function for random heterogeneous materials.
\newblock \emph{Phys. Rev. A}, 45:\penalty0 922--929, 1992.
\newblock \doi{10.1103/PhysRevA.45.922}.

\bibitem[Ebner et~al.(2014)Ebner, Chung, Garc{\'i}a, and Wood]{ebner2014}
M.~Ebner, D.-W. Chung, R.~E. Garc{\'i}a, and V.~Wood.
\newblock Tortuosity anisotropy in lithium-ion battery electrodes.
\newblock \emph{Adv. Energy Mater.}, 4:\penalty0 1301278, 2014.
\newblock \doi{10.1002/aenm.201301278}.

\bibitem[Usseglio-Viretta et~al.(2018)Usseglio-Viretta, Colclasure, Mistry, et~al.]{usseglio2018}
F.~L.~E. Usseglio-Viretta, A.~Colclasure, A.~N. Mistry, et~al.
\newblock Resolving the discrepancy in tortuosity factor estimation for {Li}-ion battery electrodes through micro-macro modeling and experiment.
\newblock \emph{J. Electrochem. Soc.}, 165:\penalty0 A3403--A3426, 2018.
\newblock \doi{10.1149/2.0731814jes}.

\bibitem[Manski(2003)]{manski2003}
C.~F. Manski.
\newblock \emph{Partial Identification of Probability Distributions}.
\newblock Springer, New York, 2003.

\bibitem[Lopez-Paz and Oquab(2017)]{lopezpaz2017}
D.~Lopez-Paz and M.~Oquab.
\newblock Revisiting classifier two-sample tests.
\newblock In \emph{International Conference on Learning Representations}, 2017.

\bibitem[Bengio et~al.(2013)Bengio, L{\'e}onard, and Courville]{bengio2013}
Y.~Bengio, N.~L{\'e}onard, and A.~Courville.
\newblock Estimating or propagating gradients through stochastic neurons for conditional computation.
\newblock \emph{arXiv:1308.3432}, 2013.

\end{thebibliography}

\section{Methods}
\subsection{Data}
Segmented X-ray tomograms of three all-solid-state composite cathodes (BM01, BM03, BM10; NMC active material, solid electrolyte, pores)
from~\cite{furat2025,dufter2026}. We use blocks of $256\times256\times192$ voxels (BM03) and $128\times128\times384$ (BM01, BM10), with the
electrode thickness along $z$ cropped to a multiple of 32.

\subsection{Effective conductivity and its gradient}
For a phase field $s\in[0,1]^{n_z\times n_x\times n_y}$, voxel conductivity is $c=\epsilon_0+(1-\epsilon_0)s$ ($\epsilon_0=10^{-4}$ during
optimisation, $10^{-7}$ for reported values). Face conductances are harmonic means of neighbouring voxels; Dirichlet potentials 1 and 0
are applied at the top and bottom faces and no-flux conditions on the lateral faces. The potential $u^*$ minimises the dissipation
\begin{equation}
E(u;c)=\sum_{\text{faces }f} g_f(c)\,(\Delta_f u)^2 ,
\end{equation}
and $M=E(u^*;c)\,n_z/(n_xn_y)$. We solve for $u^*$ with Jacobi-preconditioned conjugate gradients, warm-started across optimisation steps.
Because $u^*$ is stationary, the envelope theorem gives $\mathrm{d}M/\mathrm{d}c=\partial E/\partial c\,|_{u^*}\,n_z/(n_xn_y)$: the gradient
needs no adjoint solve, only automatic differentiation of $E$ with $u^*$ held fixed.

\subsection{Statistics and noise model}
Per 32-voxel slab in $z$, along each axis (equivalently, averaged over the sections of all three orthogonal families), we compute: L0 phase fractions; L1 boundary density per phase and axis;
L2 two-point correlations $S_{ij}(h)$ for all phase pairs, $h\in\{1,2,4,8,16,32\}$, per axis; L3 L-shaped three-point correlations
$S_{iii}$ (arms 1--8 voxels); L4 lineal-path functions for segments of 2--32 voxels. We also constrain the phase fractions of every section to lie within
the real range of section fractions (to within 0.01), and the depth profile of phase fractions in 8-layer bins. The noise scale of each statistic is the
standard error from the four lateral quadrants of the real block. A field is feasible if every standardised deviation lies inside a
Bonferroni-corrected 95\% band. For the depth profile we use a fixed scale of 0.004 per bin with the same Bonferroni band, which allows about $\pm0.015$ in phase
fraction per 8-layer bin and about $\pm0.017$ per layer when every layer is constrained.

\subsection{Optimisation}
We parametrise the field by logits $\theta\in\mathbb{R}^{3\times n_x\times n_y\times n_z}$, initialised at the real block. The forward
pass uses the hard argmax phase; gradients pass through the softmax (straight-through estimator~\cite{bengio2013}). The objective
$\mp\log M$ is active only while the field is feasible; constraints enter by an augmented Lagrangian. After 25 consecutive infeasible
steps we backtrack to the last feasible field and halve the learning rate. For resolution robustness, $M$ is evaluated on the $k^3$
majority-filtered field (straight-through), and every reported witness is re-evaluated after $k=1,3,5$ filtering against the real
block filtered with the same kernel. A run ends at its step budget or when the learning rate falls below a floor ($10^{-4}$;
continuations restart from a saved endpoint with floor $10^{-6}$). For the budget-limited twin, the fraction of voxels that differ from the
real block enters as an additional constraint, at most 3\%. Run settings for every witness are listed in Supplementary Table~S1.

\subsection{Classifier two-sample tests}
Each test trains a binary CNN classifier to separate two sources of one-hot phase images and reports held-out accuracy; 0.5 means the
sources cannot be told apart. The classifiers have three or four convolutional blocks (convolution $3\times3$, batch normalisation, ReLU,
$2\times2$ max pooling; 32--128 channels), global average pooling and a linear output, trained with Adam (learning rate $10^{-3}$,
batch 64, binary cross-entropy). Training stops early on validation accuracy (patience 6, at most 40 epochs), and the best
validation checkpoint is tested. We report five seeds.

\emph{Generator audit.} Fixed pools of $128^2$ sections (4{,}096 train, 1{,}024 validation, 1{,}024 test) per source; with 2{,}048 test images the standard error of an accuracy is at most 0.011. Generators draw
pools from separate realisations; the real tomogram is split by lateral position with a 16-pixel gap, which gives the calibrated null
(real versus real, 0.50). Coarsened variants apply $k\times k$ majority-vote pooling before classification.

\emph{Twin detectability.} Real block versus twin: $64^2$ crops of sections in all three orientations, with the same crop positions for both
classes, and a spatial split along $x$ (train $x<128$, validation $128\le x<192$, test $x\ge192$) so that the pools never share voxels (8{,}192 training, 2{,}048 validation and 2{,}048 test crop positions per class).
We also report the fraction of test crops that differ at all and the accuracy restricted to those crops.

\section*{Data availability}
The segmented tomograms are those of refs.~\cite{furat2025,dufter2026} (Giessen University data repository,
\url{https://doi.org/10.22029/jlupub-18458}).
All eleven witness structures of Tables~\ref{tab:witness} and~\ref{tab:max}, the outputs behind every reported number and the
optimisation logs will be deposited on Zenodo with a DOI on journal publication and are available from the corresponding author on request until then.

\section*{Code availability}
The bound engine (differentiable conductivity solver, statistics, optimiser), the filter-robust evaluation, the classifier two-sample
tests and all figure scripts with the exact command line of every reported run, will be released under the MIT licence on GitHub and archived on Zenodo
with a DOI on journal publication, and are available from the corresponding author on request until then.

\section*{Author contributions}
M.R. designed the study, wrote the code, ran the computations and wrote the manuscript. D.U. contributed to the interpretation and
revised the manuscript.

\section*{Acknowledgements}
We thank the authors of refs.~\cite{furat2025,dufter2026} for making the segmented tomograms publicly available. This work received no
external funding. Generative AI tools (Claude, Anthropic, used through the Cursor editor) assisted with code development, figure
scripting and language editing; all methods, computations, results and text were checked by the authors, who take full responsibility.

\section*{Competing interests}
D.U. is employed by Saab, Inc.; this work is independent research and was not funded by Saab. The authors declare no other competing interests.

\clearpage
\section*{Supplementary Information}
\addcontentsline{toc}{section}{Supplementary Information}
\setcounter{table}{0}\setcounter{figure}{0}
\renewcommand{\thetable}{S\arabic{table}}\renewcommand{\thefigure}{S\arabic{figure}}

\section*{S1. Run settings of the reported witnesses}
\begin{table}[h]\centering\small
\caption{Settings of every witness in Tables~1 and 2 of the main text. All runs: L4 statistic hierarchy, three section families, 32-voxel slabs,
depth-profile scale 0.004, Bonferroni-corrected 95\% band ($c_{95}$), initialised at the real block. $K$: number of constrained statistics.
Steps: steps taken until the run ended (step budget or learning-rate floor). $^\dagger$Stopped at floor $10^{-4}$, continued from the endpoint with floor $10^{-6}$.}
\label{tab:runs}
\begin{tabular}{llcccccl}\toprule
witness & block (voxels) & $K$ & $c_{95}$ & objective & depth bins & steps & stop \\\midrule
BM01 SE plain & $128^2\times384$ & 2340 & 4.250 & $M$ & 8 layers & 3000 & budget \\
BM10 SE robust & $128^2\times384$ & 2340 & 4.250 & $M(k{=}3)$ & 8 layers & 1450 & floor \\
BM03 SE run A & $256^2\times192$ & 1170 & 4.092 & $M(k{=}3)$ & 8 layers & 3950 & floor \\
BM03 SE run B & $256^2\times192$ & 1170 & 4.092 & $M(k{=}3)$ & 8 layers & 3500 + 700 & floor$^\dagger$ \\
BM03 SE per layer & $256^2\times192$ & 1170 & 4.092 & $M(k{=}3)$ & 1 layer & 8554 + 550 & floor$^\dagger$ \\
BM03 SE 3\% budget & $256^2\times192$ & 1170 & 4.092 & $M(k{=}3)$ & 8 layers & 9000 & budget \\
BM03 AM robust & $256^2\times192$ & 1170 & 4.092 & $M(k{=}3)$ & 8 layers & 7150 & floor \\\midrule
\multicolumn{8}{l}{\emph{maximisation (Table 2 of the main text)}}\\
BM01 SE max & $128^2\times384$ & 2340 & 4.250 & $M(k{=}3)$ & 8 layers & 9000 & budget \\
BM03 SE max & $256^2\times192$ & 1170 & 4.092 & $M(k{=}3)$ & 8 layers & 9000 & budget \\
BM03 AM max & $256^2\times192$ & 1170 & 4.092 & $M(k{=}3)$ & 8 layers & 9000 & budget \\
BM10 SE max & $128^2\times384$ & 2340 & 4.250 & $M(k{=}3)$ & 8 layers & 9000 & budget \\\bottomrule
\end{tabular}\end{table}

\section*{S2. Convergence and feasibility}
\begin{figure}[h]\centering\includegraphics[width=\linewidth]{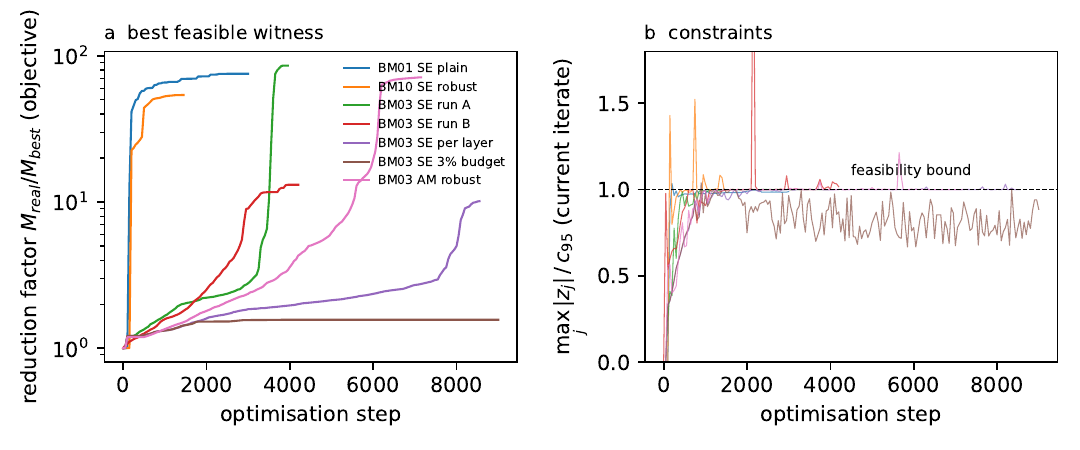}
\caption{Optimisation histories of the witnesses in Table~S1. (a) Reduction factor of the best feasible field found so far, on the
objective (unfiltered $M$ for the plain run, $M$ after $3^3$ filtering otherwise). Long plateaus followed by sharp drops (run A at
step ${\approx}3{,}500$, the per-layer twin after step 7{,}500) show that a stalled run is no evidence of a global minimum.
(b) Largest standardised deviation of the current iterate, divided by the feasibility bound. Iterates may leave the feasible set
briefly; a field is recorded as a witness only at steps where it satisfies every constraint (ratio $\le 1$, section envelope and depth profile).}
\label{fig:conv}\end{figure}

\section*{S3. Statistics and noise model in detail}
The block is divided into 32-voxel slabs along the thickness $z$. Within each slab, every statistic below is computed on the hard
one-hot phase field along each axis $x$, $y$, $z$; a statistic along an axis is the average over all sections containing that axis,
so the set covers top views and both cross-section families.
\begin{itemize}
\item L0: phase fractions of electrolyte, active material and pore.
\item L1: boundary density per phase and axis (mean absolute difference of the phase indicator between neighbouring voxels).
\item L2: two-point correlations $S_{ij}(h)$ for all six phase pairs, lags $h\in\{1,2,4,8,16,32\}$ voxels per axis (lags shorter than the axis).
\item L3: L-shaped three-point correlations $S_{iii}$ for each phase in each of the three planes, arm lengths $h\in\{1,2,4,8\}$.
\item L4: lineal-path functions $L_i(\ell)$ per phase and axis, segment lengths $\ell = 2,4,\dots,32$.
\end{itemize}
The noise scale $\sigma_j$ of each statistic is the standard error of the mean over the four lateral quadrants of the real block
(quadrant standard deviation divided by 2). A field satisfies the statistics if $|z_j| = |s_j - s_j^{\mathrm{real}}|/\sigma_j \le c_{95}$ for
all $j=1,\dots,K$, with $c_{95} = \Phi^{-1}(1-0.025/K)$ (two-sided 95\%, Bonferroni over $K$). Two further constraints apply: the phase
fraction of every individual section (all three families) must stay within the real block's range of section fractions, to within 0.01;
and the depth profile of phase fractions (8-layer bins, or every layer) must match the real one within $\Phi^{-1}(1-0.025/K_z)\times0.004$,
with $K_z$ the number of profile entries. A field is recorded as a witness only at steps where all three constraint sets hold.
Quadrant-to-quadrant variation within one tomogram is smaller than variation between electrodes or between micrographs, so these bands
are conservative: they constrain the twins more tightly than real measurement noise would.

\section*{S4. Robustness checks on the constraint set (BM03 electrolyte, plain objective)}
\begin{table}[h]\centering\small
\caption{Extremes of $M$ under variants of the constraint set, 3{,}000 steps each (real block 0.198). Neither extreme depends on the band width, and adding
statistics along the full thickness or along diagonals does not close the gap.}
\begin{tabular}{lcc}\toprule
variant & $M_{\min}$ & $M_{\max}$ \\\midrule
baseline (band $c_{95}$) & 0.0041 & -- \\
band $\times0.5$ & 0.0038 & 0.268 \\
band $\times2$ & 0.0043 & 0.270 \\
+ whole-thickness lineal path in $z$ (64--128 voxels) & 0.0042 & -- \\
+ $S_2$ and lineal path along both diagonals of all three planes (2 seeds) & 0.0037--0.0041 & 0.268 \\\bottomrule
\end{tabular}\end{table}
These plain-objective witnesses cut the electrolyte with thin seams that heal under $3^3$ majority filtering; they motivated the
resolution-robust objective used for the main results.

\end{document}